\documentclass[useAMS,usenatbib,psfig]{mn2e}

\usepackage{graphicx}
\title[Steady shocks with negative energy]
{Steady shocks around black holes produced by sub-keplerian flows
with negative energy}
\author[ D. Molteni, G. Gerardi, V. Teresi]{ D. Molteni$^1$\thanks{E-mail:
molteni@difter.unipa.it (DM)}, G. Gerardi$^1$ and V. Teresi$^{1}$ \\
$^{1}$Dipartimento di Fisica e Tecnologie Relative,
Universit$\grave{a}$
di Palermo, Viale delle Scienze, Palermo, 90128, Italy\\
}
\begin{document}
\maketitle
\label{firstpage}
\begin{abstract}
We discuss a special case of formation of axisymmetric shocks in
the accretion flow of ideal gas onto a Schwarzschild black hole:
when the total energy of the flow is negative. The result of our
analysis enlarges the parameter space for which these steady
shocks are exhibited in the accretion of gas rotating around
relativistic stellar objects. Since keplerian disks have negative
total energy, we guess that, in this energy range, the production
of the shock phenomenon might be easier than in the case of
positive energy. So our outcome reinforces the view that
sub-keplerian flows of matter may significantly affect the physics
of the high energy radiation emission from black hole candidates.
We give a simple procedure to obtain analytically the position of
the shocks. The comparison of the analytical results with the data
of 1D and 2D axisymmetric numerical simulations confirms that the
shocks form and are stable.

\end{abstract}
\begin{keywords}
accretion, accretion disks --- black hole physics ---hydrodynamics
---  instabilities
\end{keywords}
\section{Introduction}
The presence of shocks in accretion flows is a relevant fact in
the theory of accretion discs. Shocks are a very efficient
mechanism to produce high energy emission of radiation and to
induce strong acceleration of particles. Spiral shocks in
accretion disks in binary systems have been studied since long
time, theoretically by Spruit \citep{Spruit87}, Chakrabarti
\citep{Chakra90} and with both 2D and 3D numerical simulations
\citep{Sawada87,Taam91,blondin,Yukawa97}. Furthermore there is
increasing observational evidence of their significant role in
determining the accretion of gas in binary systems
\citep{Hachisu}. However the strength of this kind of shocks is
usually weak.

A different scenario for shock production has been proposed by
Chakrabarti \citep{Chakra90}. It has been widely tested with
numerical simulations \citep{Chakra93,Molteni94,Ryu96}. In this
view, an axisymmetric flow of gas, rotating with an angular
velocity less than keplerian (we will call it "sub-keplerian"
flow), and with small viscosity, can produce a steady shock. This
fact is mainly due to the deceleration of the radial fall and
consequent heating of the gas close to the centrifugal barrier.
Although the origin of such flows seems unclear there are
arguments to support the hypothesis that they really exist. In
galactic nuclei the ambient gas feeding the central black hole may
be quite well symmetric and with small rotation as shown by
observations \citep{Cohen97}. On the theoretical side, taking into
account also the role of the viscosity, a detailed study of the
requirements for their origin from a keplerian disk has been
produced \citep{Chakra96}. These flows may explain very well the
two frequencies in the QPO phenomenon in galactic black hole
candidates \citep{Chakra04}. So even if their direct proof of
existence is still pending, we think it is useful to study their
properties and characteristics.

In any case it seems to us that the phenomenon, despite of its
basic physical straightforwardness, has not been adequately
considered. Maybe some "unconscious" questions are raised against
their effective existence: 1) These shocks appear when the flow
has well defined parameters, derived from the analytical
treatment. 2) In the case of non ideal gas (i.e. if a real
cooling, an heating or conduction are included) the analytical
treatment is very difficult. 3) Even in the simple ideal gas case,
the standard procedure \citep{Chakra90} to derive a solution is
rather complicated: one has to compute the sonic point location
and then compute the boundary conditions on derivatives of the
radial speed, sound speed etc. \space {\it at the sonic point}
\space itself, leading to cumbersome formulae. 4) At the end of
the whole process the analytical solution is numerical and
specific for the set of the chosen parameters. So no simple
general formula, like the Shakura- Sunyaev disk structure
formulae, is available.

Another doubt is that the parameter space seems to be limited or
too compelling. Roughly speaking, people might say that the
sources of gas do not know the boundary conditions at the sonic
point or at infinity, they spread out their gas as they like. So,
for example, in bound systems, the total energy will be negative
and, up to now, only positive energy solutions have been discussed
and simulated. Even standard ADAF models (see \citet{Blandford}
and references therein), that are similar to the mentioned
sub-keplerian flows, require positive energies to fit
observational data. So what about negative energy flows? Here, we
explain that, even in this case, there are solutions admitting
permanent shocks. Obviously this does not mean that in every
sub-keplerian flow there is a standing shock at work. But when the
physical parameters of the accretion (essentially the rotation
amount and the thermal energy content) are appropriate, then the
shock process occurs and produces the relevant phenomena described
in previous references.

The arguments of the paper are given as follows: in section 2 we
revise the general accretion formulae for ideal gas with negative
energy, in Section 3 we present the results of numerical
simulations and in Section 4 we make our final discussion.

\section{Accretion of ideal gas with negative energy}
To explain the shock formation, let us resume the very basic
physical ingredients of the phenomenon. We assume that an
axisymmetric flow of an inviscid gas is falling from very large
distance onto a black hole and has reached a steady state regime.
We will adopt the Paczy\'nski \& Wiita potential \citep{Pac80} to
mimic the general relativistic physics. The basic physical effect
can be easily understood in terms of "classical" physics and it is
well known that the Paczy\'nski \& Wiita force reproduces many
relativistic effects with high accuracy. As we said in the
introduction, to obtain the steady state solution we may integrate
the differential equations for mass, momentum and energy
conservations. In this case one has to start the space integration
from the sonic point as explained in the work by \citet{Chakra90}.
However in the case of inviscid flow it is easy to find an
algebraic implicit solution, since (in this case) the total energy
is conserved. The Bernoulli theorem is valid and it can be
exploited to close the system of equations and to find the
solutions.
For the 2D case (with real Z extension of the disk) we assume that
the gas falls down in a condition of vertical equilibrium, with
essentially zero vertical speed.

The mass conservation equation is given by:
\begin{equation}
\label{1} \dot{M}=-4 \pi r H \rho v_r=const
\end{equation}

where H is the half thickness of the disk.\\
The vertical equilibrium hypothesis gives the well known
expression for the half disk thickness:
\begin{equation}
\label{2} H= {\frac {\sqrt {G{  M_{*}}\,r} \left( r-{r_g} \right)
{a}}{ G{  M_{*}}}}
\end{equation}

We could use a more accurate formula, but the results are
essentially unchanged while the formulae are more complicated.

According to the Bernoulli theorem the energy equation is given
by:
\begin{eqnarray}
\frac{\lambda ^2}{2r^2}+\frac{1}{2}v_r^2+\epsilon\left( r\right)
+\frac{P\left( r\right) }{\rho \left( r\right) }+\Psi \left(
r\right)  &=E
\end{eqnarray}

We will look for isentropic solutions and then we will see how to
connect solutions with different entropies by shocks. For a flow
with constant entropy $S$ and with specific heat at constant
volume $C_{v}$ we may exploit the adiabatic relation:

\begin{eqnarray}
\rho={e^{-{\frac {S}{{ C_{v}}\, \left( \gamma-1 \right) }}}}
\left( {\frac { a ^2}{\gamma}} \right) ^{\frac 1 {(\gamma -1 )} }
\end{eqnarray}

that allows us to eliminate the density $\rho$ from the equations
and to use the sound speed.

So we have two unknown quantities  $v_r$, $a$ and two equations.
Resolving for $a$ from the Bernoulli relation, using the radial
Mach number $M=-\frac{v_r}a$, and putting all terms into the
continuity equation, we have the following implicit solution for
the Mach number:
\begin{eqnarray}
\dot{M}=- 4 \pi A(r) f(M)
\end{eqnarray}

where  $f(M)$ is the Mach function which depends only on the $M$
values, given by:

\begin{eqnarray}
f(M)=q \frac{M}{\left[ (\gamma-1){M}^{2}+2 \right]
^{\frac{\gamma}{\gamma-1}}}
\end{eqnarray}

where $q=\left( \gamma-1 \right) ^{ \left( \gamma-1 \right)
^{-1}}$.

This function (we name it the "Mach" function) has a maximum at
$M_{max}$, whose value is $M_{max}=\frac{2}{\sqrt{ 2(\gamma+1)}}$

$A$ is function only of $r$; it is given by:%

\begin{eqnarray}
A( r ) = w \ {r}^{-\frac {3+\gamma}{2 (\gamma-1)}} \left[ {r}^{2}
\left( E-1/2 \,{\frac {{L}^{2}}{{r}^{2}}}+{\frac {{\it
GM}}{r-r_{{g}}}} \right)
 \right] ^{{\frac {\gamma}{\gamma-1}}} \left( r-r_{{g}} \right)
\end{eqnarray}

with $w={2}^{{\frac { \gamma}{\gamma-1}}} \ {\gamma}^{- \left(
\gamma-1 \right) ^{-1}} \left( \gamma-1 \right) { e^{-{\frac
{S}{{\it Cv}\, \left( \gamma-1 \right) }}}}{\frac { 1  }{\sqrt
{{\it GM}}}}$.

If the E value is positive this $A(r)$ function may have two
minima. The inner minimum appears when the $\lambda$ value
increases. When the total energy of the gas is negative we expect
the system to be bound and therefore the solutions describing gas
inflow into the black hole should not extend to infinity. Now the
$A(r)$ function goes asymptotically to zero at a maximum radius
and may have a relative minimum, at the radial position $r_1$,
close to the black hole if the angular momentum is large enough.
Figure 1 shows the A function for $\lambda=1.67$ and different E
values (-0.00001, -0.00003, -0.00009, -0.00027); higher lines
correspond to larger $E$ values (smaller absolute values). Figure
2 shows the A function for the same energy $E=-0.00001$ and
different, increasing, angular momentum values (1.67, $1.67 \times
1.01$ , $1.67 \times 1.02$ , $1.67 \times 1.03$); higher lines
correspond to lower $\lambda$ values. Both figures are drawn with
$S=0$ and $r$ starting from $r=1.5$ to avoid the singularity at
$r=1$. For the same $S$ value and with $\gamma=\frac {4}{3}$ we
have the value  $w=3.181980514$.

\begin{figure}
\begin{center}
\includegraphics[scale = 0.35,angle = 270.0]{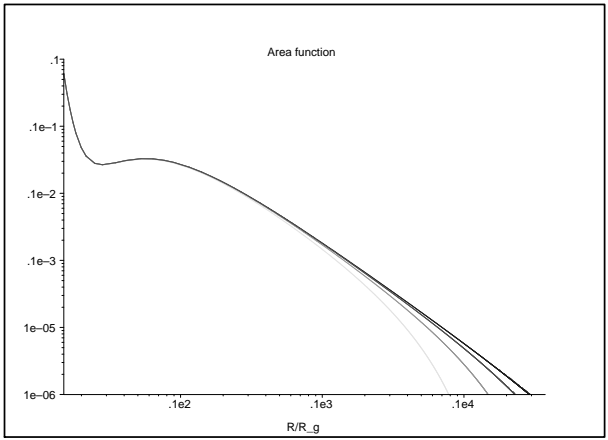}
\caption{A(r) function for different E values. Higher lines
correspond to larger $E$ values (smaller absolute values).}
\end{center}
\end{figure}

\begin{figure}
\begin{center}
\includegraphics[scale = 0.35,angle = 270.0]{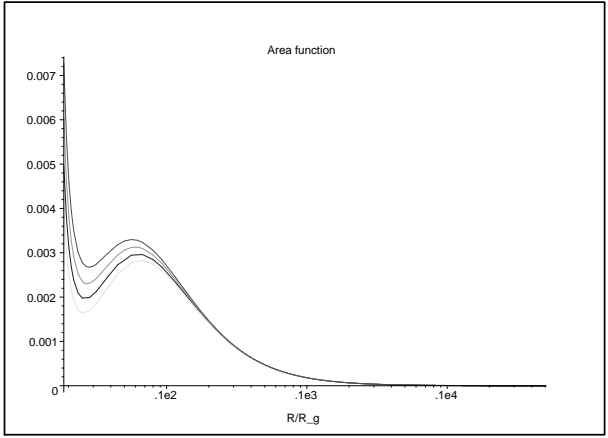}
\caption{A(r) function for different angular momentum values.
Higher lines correspond to lower $\lambda$ values.}
\end{center}
\end{figure}

The radius $r_{0}$ at which $A(r)$ goes to zero can be found
analytically; it depends only on E and $\lambda$, not on the
$\gamma$ of the gas.

Since the product of the Mach function $f(M)$ and the $A(r)$
function must be constant along the flow and since the Mach
function has a single maximum, then the minimum of $A(r)$ at $r_1$
must be coincident with the maximum of $f(M)$.

Imposing $f(M)A(r)=f(M_{max})A(r_1)$ we obtain the solutions for
$M$. Starting from the value $f(M_{max})A(r_1)$ we may obtain a
new variety of solutions just solving for M the implicit equation
$f(M)A(r)=f(M_{max})A(r_1) \xi $. The factor $\xi$ scales the
entropy of the flow. Decreasing the value of $\xi<1$ we obtain
families of solutions with the Mach number $M$ shifted to larger
values.

For any supersonic solution $M_{sup}$, by using the Hugoniot
relation we may calculate the post shock Mach value if a shock at
a generic radius value $r$ occurred:

\begin{equation}\label{}
M_{post}=\sqrt { \left( { M_{sup}}^{2}+2\, \left(\gamma-1\right)
^{-1}
 \right)  \left( 2\,{\frac {{M_{sup}}^{2}\gamma}{\gamma-1}}-1
 \right) ^{-1}}
\end{equation}

A standing shock can occur in the solution at $r_{shock}$ if the
values $M_{post}(r_{shock})$ and $M_{sub}(r_{shock})$ are equal:

\begin{equation}\label{}
M_{sub}(r_{shock})=M_{post}(r_{shock})
\end{equation}

Among the solutions with different $\xi$ values (i.e. with
different entropy values) we will find the ones that can be
connected by a shock jump with the subsonic branch. Figure 3 shows
the solutions for the Mach number with $\gamma=4/3$,
$\lambda=1.67$ and $E=-0.00001$. The inner curve corresponds to
$\xi=1$, the larger one to $\xi=0.5$.  It is clear that the
solutions can reach a maximum radius $r_{m}=21.75$: they do not
extend to infinity. For this case the shock is at
$r_{shock}=6.85$.

\begin{figure}
\begin{center}
\includegraphics[scale = 0.35,angle = 270.0]{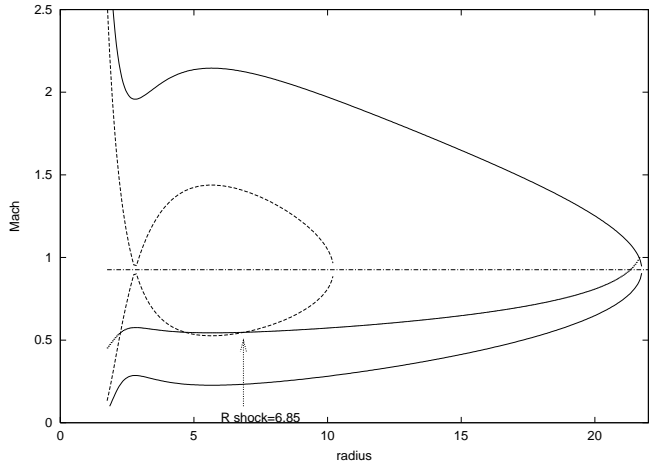}
\caption{Analytical solution with shock. Mach number versus
radius.}
\end{center}
\end{figure}

\section{Numerical simulations}
We report here the results of simulations performed using a
Smoothed Particles Hydrodynamic (SPH) code in cylindrical
coordinates developed by the authors. The SPH formulae have been
reported in previous publications \citep{Chakra93}. We stress here
that a very accurate conservation of the angular momentum of the
flow is an essential condition to obtain the shock solutions. For
the numerical simulations we use dimensionless quantities. We
adopt the speed of light as reference speed and the gravitational
radius of a Schwarzschild black hole $r_{g}=2GM_{*}/c^2$ as
reference length.

The comparison of simulation data with 2D analytical solutions
requires some care. In the analytical treatment we assumed
vertical equilibrium, but numerical simulations allow the gas
motion in the vertical direction. Therefore it is quite reasonable
that the shocks, in the simulations, will not be exactly
coincident with the theoretical predicted positions. We add, as
comment, that  we made simulations also with a fictitious damping
acting only on the $v_z$ speed. The vertically damped simulations
show a shock position in better agreement with theory.

We present here the data coming from the simulation of a 2D case
with the parameters used to obtain Figure 3. This case has the
following parameters: $E=-0.00001$, $\lambda=1.67$. The shock is
predicted to be at $r_{shock}=6.85$. We chose to inject matter at
$r_{in}=20$. With this parameters the radial speed and the sound
speed at $r_{in}$ are $v_{in}=-0.09606$, $a_{in}=0.07790$
respectively. The spatial resolution of the SPH particle is
$h=0.2$. The steady state particle number is $N_{p}=63000$.

Figure 4 shows the particle distribution in the upper half of the
domain. The shock of the simulation is at $r_{shock-SPH}=8.7$.
\begin{figure}
\begin{center}
\includegraphics[scale = 0.35,angle = 270.0]{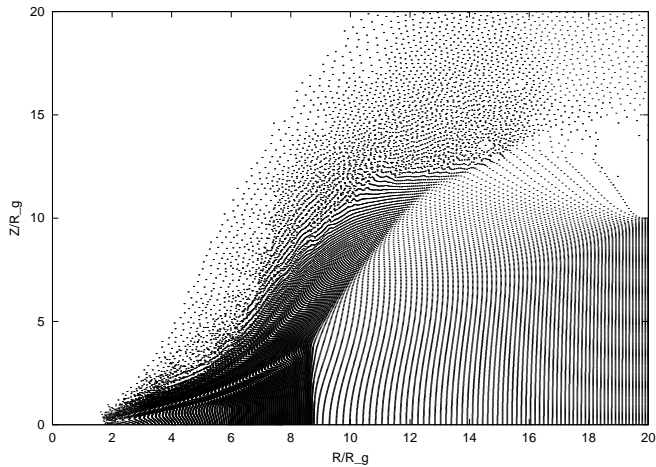}
\caption{2D simulation: Particle positions.}
\end{center}
\end{figure}

Taking into account our previous remarks we may say that the shock
position of the 2D simulation is in good agreement with the
theory. We also report in Figure 5 the Mach value versus the
radial distance of a 1D simulation. The physical parameters for
the 1D simulation are $E=-0.003$, $\lambda=1.85$, $\gamma=5/3$;
the numerical parameters are $r_{in}=75.26$, $v_{in}=-0.04719$,
$a_{in}=0.03931$,  $h=0.125$. In this case, since there is no
vertical motion, the agreement between theory and simulation must
be exact, and effectively this occurs. For clarity, in the figure
we plotted the theoretical Mach value plus a small shift($0.05$).
The shock is at the predicted position of $r_{shock}=12.8$. The
whole shape of the curve profile is very well reproduced. Note
that the 1D solution is obtained in the same way of the 2D case,
but the functions $A(r)$ and $f(M)$ are different.
\citep{Molteni99}.

\begin{figure}
\begin{center}
\includegraphics[scale = 0.35,angle = 270.0]{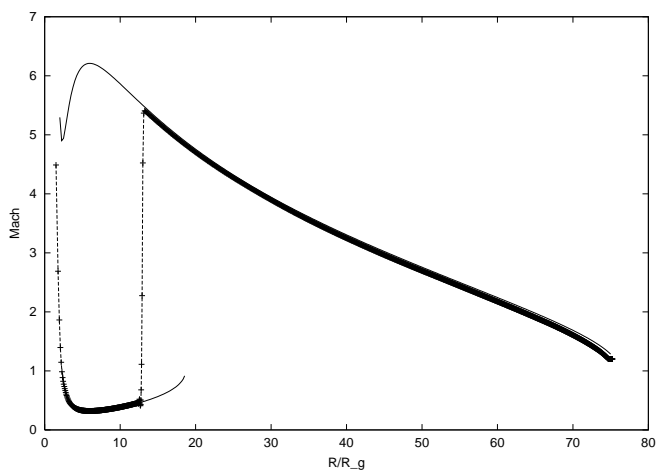}
\caption{1D simulation: Mach versus radial distance. The
continuous line is the theoretical one.}
\end{center}
\end{figure}

\section{Conclusions}
With our analysis we reinforce the view that the existence of
permanent shocks around black holes is an important aspect of the
physics of accretion flows. It has been suggested that the QPO
oscillations of the X-ray luminosity of galactic black hole
candidates may arise from shock oscillations \citep{Chakra04}.Jets
of out-flowing matter may arise also from the high energetic post
shock gas driven by radiation \citep{chatto}. These shocks produce
around the central object a bulge of hot gas comptonizing the low
energy thermal emission from a cold keplerian disk. We remark that
shocks produce the hot corona in a quite natural way, while the
corona outcome or the ADAF solutions originating from keplerian
disks have been shown by Molteni et al. to be very unlikely
\citep{Molteni01adaf}.

We stress that these flows are essentially axisymmetric. So this
situation is different from the one in which the gas flow, in a
binary system, is captured from the wind originated by the normal
star. Steady shocks of this kind cannot be produced in this case
since the flow has low angular momentum, but has {\it
non-axisymmetric} configuration.

Obviously a crucial point is the question of the origin of such
symmetric sub-keplerian flows. It seems to us that this problem is
not critical for black holes in the centers of galaxies or AGN,
since in this case the star environment is nearly spherically
symmetric and with low global rotation \citep{Cohen97}. In the
case of black holes in binary systems we suggest that accretion
may occur not only from the inner lagrangian point, but also by  a
small amount of ambient gas, possessing low rotation, in the
potential well of the binary. Time dependent 3D numerical
simulations \citep{Molteni01bin} and recent observations support
this hypothesis, see \citet{smith}. Furthermore it was suggested
by Carrol \citep{carrol} that the Poynting-Roberston effect can be
efficient also to drive matter from keplerian disks even in
accretion on White Dwarfs. In the case of accretion on weakly
magnetized neutron stars it has been shown \citep{miller} that if
the luminosity of the star is  $\sim 0.2 L_{E}$  ( $L_{E}$ is the
Eddington luminosity) a substantial fraction of the accreting
matter can lose its angular momentum. We also think that, for the
binary black hole case, the disk self-illumination may produce
similar effect decreasing the angular momentum of the outer layers
of the disks. This process would be similar to the accretion
driven by radiation suggested by Ballantyne \citep{ballan} to
occur in neutron star luminosity bursts. We add a trivial
consideration: even if the inflow parameters do not correspond to
the exact analytical ones, but are close to them, steady shocks
may be still formed.

If requested, the authors can offer the software to reproduce both
the analytical and the numerical results.

\end{document}